\documentstyle[11pt,aaspp]{article}

\received{23 September 1994}
\accepted{18 November 1994}

\slugcomment{Accepted for publication in The Astrophysical Journal Letters}

\begin{document}

\title{MILLIMETER OBSERVATIONS OF A COMPLETE SAMPLE \\
OF IRAS GALAXIES: DUST EMISSION \\
AND ABSORPTION IN SPIRALS \altaffilmark{1}}


\author{Alberto Franceschini\altaffilmark{2}}
\affil{Osservatorio Astronomico di Padova, Padova, Italy}

\and

\author{Paola Andreani\altaffilmark{3}}
\affil{Dipartimento di Astronomia di Padova, Padova, Italy}

\altaffiltext{1}{Based on observations collected at the ESO-Swedish
SEST 15m telescope (La Silla, Chile).}
\altaffiltext{2}{Osservatorio Astronomico, Vicolo Osservatorio 5, I-35122,
Padova, Italy. E-mail: Franceschini@astrpd.pd.astro.it}
\altaffiltext{3}{Dipartimento di Astronomia, Vicolo Osservatorio 5, I-35122,
Padova, Italy. E-mail: Andreani@astrpd.pd.astro.it}

\begin{abstract}

We report on observations performed at 1.25 $mm$ of a southern galaxy sample,
selected from the IRAS PSC and complete to $S_{60}=2\ Jy$. We detected 18
sources and set significant limits on 10 further objects. We use these data
to discuss the spatial distribution of cold dust, the broad-band far-IR/mm
spectra, the overall amount of dust, the gas-to-dust mass ratio, the dust
optical depth, and the overall extinction, for such a representative galaxy
sample. These results are also supported by a successful comparison with
values inferred for the Galaxy. Because of the favourable
observational setup, selection wavelength and completeness, we believe these
data provide an unbiased view of dust properties in spiral galaxies.

\end{abstract}

\keywords{ galaxies: ISM - galaxies: photometry - galaxies:
spirals, starburst - ISM: dust, extinction - infrared: ISM: continuum

\section{Introduction}

Observations of diffuse dust in galaxies have a profound impact on our
understanding of some basic questions about their structure and past
history. IRAS survey data by themselves, unfortunately, do not
completely characterize the dust content and emission properties.
Dust colder than 20 K, whose existence is currently
matter of so much controversial discussions, can only be sampled at
$\lambda>100 \mu m$ (Chini et al., 1986).

We report here on a 1.25 $mm$ continuum survey of a complete sample of
IRAS galaxies performed with the SEST
telescope. The latter was chosen because it provided the best compromise
between detector sensitivity and spatial resolution. The 24" FWHM of the
SEST beam at 1.25 mm and the average optical extent of the sample
galaxies match favourably, so that the beam-aperture corrections needed
to compare with the IRAS survey data are not as severe as for other
observational setups.

Observations at such long wavelengths ensure that all possible dust
components with temperatures down to the fundamental limit, set by the 3 K
background radiation, are properly sampled. The use of a complete
flux-limited sample ensures a minimal exposure to the effects of bias.
Finally, adoption of a far-IR selected sample, rather than of an optical one,
makes us confident that the whole phenomenology of dust effects in galaxies
is properly explored.

\section{Observations and data analysis}

The target sample has been selected from the IRAS PSC and is complete to
S$_{60 \mu m} \geq $2 Jy.
It consists of 29 galaxies with morphological types
from S0/a through Scd, within the sky region $21^h < \alpha (1950)  < 5^h$ and
$-22.5^\circ <\delta<-26.5^\circ$.
Distances (mostly from distance indicators, others derived using
$H_0=75$) cover the range from $d\simeq 20$ to $\simeq 200$ Mpc.
Half-optical-light diameters $A_e$ are typically within $A_e\simeq 10 "$
to $\simeq 40"$, 3 exceptionally close galaxies having $A_e\leq 100"$.
The volume test ($<V/V_{max}>=0.45\pm 0.06$) ensures the sample completeness.

The observations have been done at 230 GHz during September 1990 and 1991 with
the MPIfR $^3$He-cooled bolometer (Kreysa 1990) at the SEST 15m telescope.
A three-beam modulation is achieved by chopping ON-OFF the source with
a beam separation of $70"$. Pointing accuracy was most of the time close to
$2"$. The overall flux accuracy is estimated to be 10 to 30\%.

\subsection{Beam-aperture corrections: the size of galaxies at
$\lambda=1.25\ mm$}

For the 4 optically most extended sources we have performed a rough mapping,
which allowed us to reduce the corresponding large aperture corrections.
More generally, our adopted procedure was to test various
different hypotheses about the distribution of the millimetric emission
(see Andreani \& Franceschini, 1992):
{\it (a)} that the mm source is a point-source; {\it (b)} that its radial
distribution is an exponential with scale-length $\alpha_{\rm mm}=1/3\
\alpha_o$ ($\alpha_o$ is the B-band scale-length); and {\it (c)} that the
distribution closely follows that of the optical light ($\alpha_{\rm
mm}=\alpha_o$).

We show in Figure 1 a plot of the ratios of the 1.25 mm to the 60 and
100 $\mu m$ fluxes
versus galaxy distance and diameter, with $S_{1.25}$ corrected
according to hypothesis {\it (c)}. We see that the latter implies rather
significant dependences of the average flux ratios, which are not expected
(note that all flux measurements have already been K-corrected).
We are led to conclude that the hypothesis of a comparable distribution of
cold dust and starlight, and even worse that assuming dust present to large
radial distances, appear to be inconsistent with our data. Instead, the case
for a higher concentration of dust, with $\alpha_{\rm mm} = \alpha_o/3$, is
supported by our observations.

\subsection{The average galaxy spectrum from 20 to 1300 $\mu m$}

Figure 2 shows the average far-IR/millimeter spectrum for
28 galaxies in our sample. All fluxes have been color- and K-corrected, taking
into account the overall system response, while $mm$ fluxes have been corrected
assuming $\alpha_{\rm mm} = \alpha_0/3$.
The average 1.25 mm to 100 $\mu m$ flux ratio turns out to be $\simeq
(3.1\pm 0.5)\ 10^{-3}$. Beam corrections following hypotheses (a)
and (c) above would bring this ratio to values of $\simeq
(2.2\pm 0.5)\ 10^{-3}$ and $\simeq (6.2\pm 0.5)\ 10^{-3}$, respectively.

The main conclusion that can be drawn from Fig.2 is that there is no evidence
in our data of major spectral components of dust emission peaking at
$\lambda > 100\ \mu m$, as previously suggested by some authors.

\section{The dust model}

Our adopted scheme of galactic dust
involves a "cirrus" component, including cold dust and small transiently-heated
hot grains, and warm dust in starbursting regions (Rowan-Robinson,
1986, 1992, hereafter RR86, RR92; Mazzei et al., 1992).

The "cirrus" component has been modelled following in detail the recipes
by RR86 and RR92. The grain mixture (nine grain types in total) and
properties have been optimized by RR92 to reproduce all basic
observables (e.g. the extinction law) over the entire spectral range from
1 mm to 0.1 $\mu m$.

For the warm dust in star-forming regions we have followed the
approach of Xu and De Zotti (1989) and Conte (1993).
The starforming region is modelled as a volume uniformly filled with
dust and radiation, and is assumed to be optically thin at least
for $\lambda > 20\ \mu m$. The same grain mixture as for the "cirrus"
component is adopted, but very small grains are assumed to be destroyed
by the intense radiation field.

Three parameters, i.e. the average radiation field intensity $\chi_c$ in
the cold "cirrus", the intensity illuminating the warm dust in starburst
regions ($\chi_w$) and the light fraction $f_w$ at 100 $\mu m$ contributed by
warm dust, fully describe the dust model.

The temperatures for all grain types of both dust components are uniquely
determined by the parameters $\chi_c$ and $\chi_w$, and obtained by solving
the usual energy balance equation. The fitting procedure assumes that the
1.25 $mm$ flux is dominated by cold "cirrus" dust, which basically defines
$\chi_c$. The 60 and 25 $\mu m$ data mostly define the other two parameters.

\section{Results}

The adopted dust model turns out to be particularly successful in
reproducing the observed far-IR/mm broad-band spectra for all the sample
galaxies.

The dashed line in Fig. 2 corresponds to the "cirrus" component fitted
to the average galaxy spectrum, the dot-dashed to the average starburst
component. We report in Table 1 various details on the physical
parameters (temperature, mass) of the dust components for this best-fit
model. Note in particular that
the warm "starburst" component is found to contribute only a few
\% of the total dust mass.

For thirteen galaxies, that we name {\it "cirrus"-dominated}, the
contribution of warm dust appears to be negligible.
Fifteen objects require, instead, high values of the warm dust
fraction ($f_w\geq 0.4$), and their spectrum at $\lambda < 100\ \mu m$ is
dominated by warm dust in the starburst component (the {\it
starburst-dominated} galaxies).

An overall view of dust effects in our IR
galaxies is given in Figure 3. The observed optical depth $\tau_B$
has been estimated from the total dust mass divided by the projected area
within one third of the optical radius, where we estimate dust is confined.
$\tau_B$, which measures the amount of dust {\it available} to absorb
optical light, is compared in Fig. 3 with a quantity $A_B$ measuring the
overall {\it actual} effect of extinction. $A_B$ has been
estimated from the logarithmic ratio of the bolometric optical-UV
luminosity ($L_O$) to the bolometric far-IR light ($L_{FIR}$).
The two galaxy classes that we have defined appear significantly
segregated over this plane, the inactive "cirrus"-dominated objects being
confined to low values of dust optical depth ($\tau_B \leq 1$) and low
extinction ($A_B \simeq 0.5$).

The active star-forming galaxies, on the contrary, are spread over much
larger values in both axes: appreciable amounts of dust and extinction seem
to characterize only these objects, with typical $\tau_B$ of 1
to 10. However, only for a minority of these objects (5/16), extinction
values significantly higher than 1 are indicated.

Finally, we report in Figure 4 a plot of the gas to dust
ratio versus bolometric luminosity ($L_{bol}=L_O + L_{FIR}$).
Gas masses (including both HI and molecular gas
estimated from CO observations) are from Andreani et al. (1994).
The median $\L_{bol}\sim 5\ 10^{10}\ L_{\odot}$ for the inactive objects
is a factor 2 lower than the corresponding value of starbursts, consistent
with our inference that the latter are characterized by an enhanced
SFR. Second, the median gas/dust ratio for the two classes also differs
($<M_g/M_d> \simeq 300$ for the starbursts and $\simeq 1000$ for the
inactive population). Finally, there is an apparent trend of $<M_g/M_d>$
to decrease with $L_{bol}$ for the inactive galaxy population.

The positions that our Galaxy would occupy in both Fig. 3 and 4
have been determined by applying our adopted dust model to
the galactic integrated spectral emissivity. Its position, well within the
region occupied by the inactive population in Fig. 3,
is consistent with our finding that the Galaxy's far-IR spectrum
is dominated by "cirrus" emission.

\section{Discussion}

A fruitful combination of good sensitivity at long wavelegths, a fairly large
beam aperture and small angular size of the target objects, have allowed us to
explore the dust content in galaxies down to the coldest possible
temperatures and over most of the optical galaxy extent.
Although our inferences on the mm size of galaxies have only a statistical
sense, our results
(Fig. 1) appear inconsistent with the assumptions that cold dust has a
scale-length comparable to or larger than that of starlight. They rather
suggest that the $mm$ scale-lenght is 1/3 or so of the optical one.
This may reflect the observed increase in metallicity towards the inner
regions of galaxies (e.g. Sodrowski et al., 1994) and may be consistent
with the idea of an enhanced past stellar activity there.

The observed millimetric emissivity of these galaxies seems also inconsistent
with large spectral components appearing longward of 100 $\mu m$.
The bolometric corrections to the far-IR IRAS data needed to estimate
the global dust emissivity appear to be rather small.

A published model (RR92) reproduces very accurately the observed
broad-band spectra. A simple color criterion, supported by model
predictions, exploiting the millimetric fluxes and IRAS data, allows to
classify these IR galaxies according to the rate of star-formation.

Model inferences on the dust mass imply B-band optical depths within one
third of the optical radius spanning a large range of values ($\tau_B=0.1$ to
10, see Fig. 3). The largest values of dust mass, optical depth and the
dust/gas mass ratio are shared by objects characterized by a starbursting
activity ($\tau_B>1$).
The inactive "cirrus"-dominated galaxies, which are the typical population
that bright optical samples select, seem almost unaffected by dust. These
results are confirmed by a comparison with the {\it actual} $A_B$ extinction
inferred from the observed ratio of optical to far-IR bolometric emissions.
Relatively high values of $A_B$ ($>1$) are displayed only by a minority of the
starbursting galaxies (5 over 29 objects of the total sample). Note,
however, that the large values of $A_B$ in these objects are likely to
reflect more the IR emission from starbursting regions than
extinction really affecting the whole galaxy body.

Our data efficiently constrain the presence of cold dust in galaxies.
In our model, dust at $T\simeq 7$ to $\simeq 20 K$ is responsible for most
of the millimetric emission (see Table 1). The corresponding dust mass
evaluation then accounts also for very cold dust components in the ISM, whose
mass fraction however appears to be modest. To get a stronger constraint, we
have also considered the possible existence of a dust component illuminated
by a very dim radiation field ($\chi = 0.5 \chi_{\odot}$), with a temperature
distribution ranging from 4 to 18 K. Our data imply that
no more than $10^7\ M_{\odot}$ of such
cold dust may exist in our average galaxy: this, added to the other two
components, would only bring to a 40\% increase of the overall estimated
dust mass with respect to our previously discussed best-fit.
Altogether, dust of any kind does not seem to play a major role in
extinguishing the stellar emission.

There are some intriguing effects to notice in Fig. 4. The starbursting
activity seems to imply a heavier dust enrichment of the ISM with respect to
inactive spirals: $M_g/M_d$ values are typically $\simeq 1000$ for the latter,
while ranging from 100 to 1000 for starbursts. The ongoing starburst
may well have reprocessed the ISM in these objects significantly longer
or more efficiently than inactive galaxies did.

Finally, there seems to be a trend favouring higher values of $M_g/M_d$ in
lower luminosity {\it normal inactive} spirals. This may parallel a similar
trend of higher metallicities at higher masses inferred for early-type
galaxies.

A comparison with results obtained for the Galaxy seems to support our
conclusions. Its observed spectral emissivity strongly supports our adopted
dust model. Its inferred overall extinction and dust column density are as
expected (Fig. 3). The somewhat low value of $M_g/M_d$ observed in Fig. 4
with respect to those of non-starbursting galaxies of the same luminosity may
simply reflect the fact that the galactic value refers to the inner dusty
regions, whereas those of galaxies in our sample include also the
contributions of gas in the outer dust-poor environments.

\acknowledgments
We are indebted to E. Kreysa, R. Chini, L. Danese and L. Toffolatti for many
fruitful discussions, and to the SEST staff for assistance.

\begin{thebibliography}{}

\bibitem\reference  Andreani P. \& Franceschini A. 1992, {\astap} 260, 89
\bibitem\reference  Andreani P., Casoli F. and Gerin M., 1994, {\astap}
submitted
\bibitem\reference  Chini R., Kreysa E., Kr\" ugel E., Mezger P.G., 1986,
{\astap} 166, L8
\bibitem\reference Conte N., 1993, Ph.D. Thesis, Padova University
\bibitem\reference Disney, M., Davies, J., and Phillips, S., 1989: {\mnras}
239, 939
\bibitem\reference  Kreysa E., 1990, in {\it From Ground-Based to Space-Born
Sub-mm Astronomy}, ESA SP-314, p.265-270
\bibitem\reference Mazzei, P., Xu, C., and De Zotti, G., 1992, {\astap} 256, 45
\bibitem\reference Rowan-Robinson, M., 1986: {\mnras} 219, 737 (RR86)
\bibitem\reference Rowan-Robinson, M., 1992: {\mnras} 258, 787 (RR92)
\bibitem\reference  Sodroski T.J., et al., 1994, {\apj} 428, 638
\bibitem\reference Xu, C., and De Zotti, G., 1989: {\astap} 225, 12.

\clearpage

\begin{figure}
\caption{Millimetric to far-IR flux ratios versus galaxy distance (left-hand
panels) and apparent diameter (right-hand).
The millimeter fluxes $S_{1.25}$ have been corrected for beam-aperture
assuming a mm scale-length equal to that of the optical light.
The two lines in each panel correspond to the $\pm 1\ \sigma$
regression fits obtained with a "survival analysis" technique. The
corresponding values of the correlation coefficient are also reported.
}
\end{figure}

\begin{figure}
\caption{Average far-IR/mm broad-band spectrum for our sample galaxies,
with the corresponding "cirrus" (dashed line) and starburst (dot-dash)
contributions. The corresponding best-fit spectral
parameters are $\chi_c=11$ (normalized the the intensity in the solar
neighbouroud), $\chi_w=120$ and $f_w=0.3$.
The continuous line is the summed contribution. Values of the physical
parameters for this dust model are detailed in Table 1.
}
\end{figure}

\begin{figure}
\caption{Observed optical depth of dust $\tau_B$ versus the overall
extinction $A_B$. $\tau_B$ is averaged within one third of the B-band
Holmberg radius (the deprojected face-on $\tau_B$
scales from this value with the galaxy axial ratio {\it b/a} squared).
$A_B$ is estimated by comparing the bolometric outputs
in the optical and far-IR. Open and filled squares refer the the
inactive "cirrus"-dominated and to the starburting objects. The solar
symbol marks values estimated for the Galaxy, by applying our dust model
to its global spectral emissivity.
The predicted dependences for a screen, a slab, and a sandwich (with zero
scale-height of dust) model are shown for comparison (Disney et al., 1989).
}
\end{figure}

\begin{figure}
\caption{The gas to dust mass ratio versus bolometric optical/mm luminosity
(symbols as in Fig. 3). The dotted
line is an eye-fit to the locii of the "cirrus" galaxies.
}
\end{figure}

\end{document}